\documentclass[aps,prl,twocolumn,showpacs,groupedaddress]{revtex4}  
\usepackage{graphicx}  
\usepackage{dcolumn}   
\usepackage{bm}        
\usepackage{amssymb}   

\newcommand{\Esum}{E_{{\rm sum}}}

\begin{document}

\title{Measurement of the Double Beta Decay Half-life
of $\boldmath ^{150}$Nd \unboldmath and Search for Neutrinoless Decay Modes
with the NEMO-3 Detector}

\author{J.~Argyriades$^{1}$}
\author{R.~Arnold$^{2}$}
\author{C.~Augier$^{1}$}
\author{J.~Baker$^{3}$}
\author{A.S.~Barabash$^{4}$}
\author{A.~Basharina-Freshville$^{5}$}
\author{M.~Bongrand$^{1}$}
\author{G.~Broudin$^{6,7}$}
\author{V.~Brudanin$^{8}$}
\author{A.J.~Caffrey$^{3}$}
\author{E.~Chauveau$^{6,7}$}
\author{Z.~Daraktchieva$^{5}$}
\author{D.~Durand$^{9}$}
\author{V.~Egorov$^{8}$}
\author{N.~Fatemi-Ghomi$^{10}$}
\author{R.~Flack$^{5}$}
\author{Ph.~Hubert$^{6,7}$}
\author{J.~Jerie$^{13}$}
\author{S.~Jullian$^{1}$}
\author{M.~Kauer$^{5}$}
\author{S.~King$^{5}$}
\author{A.~Klimenko$^{8}$}
\author{O.~Kochetov$^{8}$}
\author{S.I.~Konovalov$^{4}$}
\author{V.~Kovalenko$^{8}$}
\author{D.~Lalanne$^{1}$}
\author{T.~Lamhamdi$^{11}$}
\author{K.~Lang$^{12}$}
\author{Y.~Lemi\`{e}re$^{9}$}
\author{C.~Longuemare$^{9}$}
\author{G.~Lutter$^{6,7}$}
\author{Ch.~Marquet$^{6,7}$}
\author{J.~Martin-Albo$^{14}$}
\author{F.~Mauger$^{9}$}
\author{A.~Nachab$^{6,7}$}
\author{I.~Nasteva$^{10}$}
\author{I.~Nemchenok$^{8}$}
\author{F.~Nova$^{15}$}
\author{P.~Novella$^{14}$}
\author{H.~Ohsumi$^{16}$}
\author{R.B.~Pahlka$^{12}$}
\author{F.~Perrot$^{6,7}$}
\author{F.~Piquemal$^{6,7}$}
\author{J.L.~Reyss$^{17}$}
\author{J.S.~Ricol$^{6,7}$}
\author{R.~Saakyan$^{5}$}
\author{X.~Sarazin$^{1}$}
\author{L.~Simard$^{1}$}
\author{F.~\v{S}imkovic$^{18}$}
\author{Yu.~Shitov$^{8}$}
\author{A.~Smolnikov$^{8}$}
\author{S.~Snow$^{10}$}
\author{S.~S\"{o}ldner-Rembold$^{10}$}
\author{I.~\v{S}tekl$^{13}$}
\author{J.~Suhonen$^{19}$}
\author{C.S.~Sutton$^{20}$}
\author{G.~Szklarz$^{1}$}
\author{J.~Thomas$^{5}$}
\author{V.~Timkin$^{8}$}
\author{V.~Tretyak$^{8}$}
\author{V.~Umatov$^{4}$}
\author{L.~V\'{a}la$^{13}$}
\author{I.~Vanyushin$^{4}$}
\author{V.~Vasiliev$^{5}$}
\author{V.~Vorobel$^{21}$}
\author{Ts.~Vylov$^{8}$}

\affiliation{\vspace{0.1 in}(The NEMO Collaboration)\vspace{0.1 in}}
\affiliation{$^{1}$LAL, Universit\'e Paris-Sud 11,  CNRS/IN2P3, Orsay, France}
\affiliation{$^{2}$IPHC, Universit\'e de Strasbourg, CNRS/IN2P3, F-67037 Strasbourg, France}
\affiliation{$^{3}$INL, Idaho Falls, Idaho 83415, USA}
\affiliation{$^{4}$Institute of Theoretical and Experimental Physics, 117259 Moscow, Russia}
\affiliation{$^{5}$University College London, WC1E 6BT London, United Kingdom}
\affiliation{$^{6}$Universit\'e de Bordeaux, 
Centre d'Etudes Nucl\'eaires de Bordeaux Gradignan, UMR 5797, 
 F-33175 Gradignan, France}
\affiliation{$^{7}$CNRS/IN2P3,  
Centre d'Etudes Nucl\'eaires de Bordeaux Gradignan, UMR 5797, 
 F-33175 Gradignan, France}
\affiliation{$^{8}$Joint Institute for Nuclear Research, 141980 Dubna, Russia}             
\affiliation{$^{9}$LPC Caen, ENSICAEN, Universit\'e de Caen, Caen, France}
\affiliation{$^{10}$University of Manchester, M13 9PL Manchester,  United Kingdom}
\affiliation{$^{11}$USMBA, Fes, Morocco}
\affiliation{$^{12}$University of Texas at Austin, Austin, Texas 78712-0264, USA}
\affiliation{$^{13}$IEAP, Czech Technical University in Prague,  CZ-12800 Prague, Czech Republic}

\affiliation{$^{14}$IFIC, CSIC - Universidad de Valencia, Valencia, Spain}
\affiliation{$^{15}$Universitat Aut\`onoma de Barcelona, Spain}
\affiliation{$^{16}$Saga University, Saga 840-8502, Japan}
\affiliation{$^{17}$LSCE, CNRS, F-91190 Gif-sur-Yvette, France}
\affiliation{$^{18}$FMFI, Comenius University, SK-842 48 Bratislava, Slovakia}
\affiliation{$^{19}$Jyv\"{a}skyl\"{a} University,  40351 Jyv\"{a}skyl\"{a}, Finland}
\affiliation{$^{20}$MHC, South Hadley, Massachusetts 01075, USA}
\affiliation{$^{21}$Charles University, Prague, Czech Republic}

\date{\today}

\begin{abstract}
The half-life for double beta decay of $^{150}$Nd has been measured
by the NEMO-3 experiment at the Modane Underground
Laboratory. Using 924.7 days of data recorded with $36.55$~g of $^{150}$Nd
the half-life for $2\nu\beta\beta$ decay is measured to be
$T_{1/2}^{2\nu} = (9.11^{+0.25}_{-0.22}(\rm stat.) \pm 0.63~(\rm syst.))
\times 10^{18}$~years.
The observed limit on the half-life for neutrinoless double beta decay
is found to be $T_{1/2}^{0\nu}>1.8 \times 10^{22}$~years 
at $90\%$ Confidence Level.
This translates into a limit on the effective Majorana neutrino mass of 
$\langle m_{\nu} \rangle<4.0-6.3$~eV if the nuclear deformation
is taken into account. We also set limits on models involving Majoron
emission, right-handed currents and transitions to excited states.
\end{abstract}

\maketitle 

Experiments studying atmospheric, solar, reactor and accelerator
neutrinos have established the existence of neutrino 
oscillations as a direct evidence for a non-zero neutrino mass.
These experiments, however, cannot distinguish between 
Dirac or Majorana neutrinos. They also provide no
information on the absolute neutrino mass scale, since oscillations
experiments measure the square of the mass difference between
neutrino states. 
The half-life of neutrinoless double beta decay ($0\nu\beta\beta$) is
inversely proportional to the effective Majorana neutrino mass squared, $\langle m_{\nu}
\rangle^2$. Observation of this process would therefore directly constrain
the neutrino mass scale and would be unambiguous evidence for
the Majorana nature of neutrinos. The $0\nu\beta\beta$ process
also violates lepton number
and is therefore a direct probe for physics beyond the standard model 
of particle physics.

The search for neutrinoless double beta decay of neodymium-150 ($^{150}$Nd)
using the NEMO-3 detector is of special interest since $^{150}$Nd is
a candidate isotope for  SuperNEMO~\cite{bib-super}, 
a next generation double beta decay experiment based on the NEMO-3 concept,
and the \mbox{SNO$++$} experiment at SNOLAB~\cite{bib-sno}.
Its main advantages are the high $Q_{\beta\beta}$ value
for double beta decay, $Q_{\beta\beta}=3.368$~MeV, which lies
above the typical energies for many
background sources, and the large phase space factor. 
The $2\nu\beta\beta$ half-life of $^{150}$Nd has
previously been measured using a Time Projection 
Chamber~\cite{bib-tpc,bib-silva}.

The NEMO-3 experiment has been taking data since 2003 in the Modane 
Underground Laboratory (LSM) located in the Fr\'ejus tunnel at a 
depth of 4800~m water equivalent. The detector has a cylindrical 
shape with 20 sectors that contain different isotopes in the form
of thin foils with a total surface of about 20~m$^2$~\cite{bib-tdr}. 
In addition
to $\sim$7~kg of $^{100}$Mo and $\sim$1~kg of $^{82}$Se, the detector
contains smaller amounts of other isotopes.
The neodymium source foil is composed of 
Nd$_{2}$O$_{3}$ with an enrichement of $(91\pm0.5) \%$, corresponding to 
a $^{150}$Nd mass of $36.55 \pm 0.10$~g.
On each side of the foils is a $\sim$50~cm wide tracking volume 
comprising a total of 6180 drift cells operated in Geiger mode with
helium as drift gas.
A $25$~Gauss magnetic field created by a solenoid 
provides charge identification. 
The calorimeter consists of 1940 plastic
scintillators coupled to low radioactivity photomultipliers. 
For 1~MeV electrons the energy resolution (FWHM) ranges
from $14.1\%$ to $17.7\%$ and the timing resolution is 250~ps.
A cylindrical coordinate system $(r,z,\phi)$ is used with the $z$ axis
pointing upwards.

The data used in this Letter have been recorded between February 2003 and
December 2006, corresponding to $924.7$ days of data taking.
Signal and background events are generated with
the {\sc genbb} generator and simulated using a {\sc geant}-based 
simulation~\cite{bib-geant}
of the detector. All Monte Carlo events are processed 
by the same reconstruction
programs as the data. A detailed description of the analysis
can be found in~\cite{bib-nasim}.

The $2\nu\beta\beta$ events are expected to have two electrons
in the final state and they are therefore selected by requiring two tracks with
a curvature consistent with a negative charge. 
Each track has to be matched to a separate energy deposit in the calorimeter 
greater than $0.2$~MeV.
The $z$ component of the distance between the intersections of each track 
with the plane of the foil
should be less than 4~cm and the transverse component less than 2~cm. Both
tracks must originate from the first layer of the Geiger cells.
The scintillator measurement of the time-of-flight (TOF) for both electron candidates
must be consistent with the hypothesis that the event originates from the 
source foil. After this selection $2789$ events remain.

The background sources are divided into two categories, depending on their
origin~\cite{bib-bg}. Background events originating from the radioactive impurities in
the source foils are called internal background.
$\beta$ emitters can produce $\beta\beta$-like events
through three mechanisms: (i) $\beta$ decay 
accompanied by electron conversion;
(ii) M\o ller scattering of a $\beta$ decay electron, and (iii) 
$\beta$ decay to
an excited state followed by Compton scattering of the de-excitation photon.
Another internal background source is $^{207}$Bi, most likely present due to
a contamination of the $^{150}$Nd source foil during production. This isotope
decays to excited $^{207}$Pb through electron capture and 
$^{207}$Pb subsequently emits two electrons via electron conversion.

The second source of background is called external and is caused by 
electrons or photons anywhere outside the source 
foils. The main source of external background is radon.
Radon decays to $^{214}$Bi via two $\alpha$ decays and
one $\beta^-$ decay. The decay of $^{214}$Bi to 
$^{214}$Po is generally accompanied by one electron and several photons, which
can mimic a $\beta\beta$ event through conversions. A radon purification
facility was installed about half-way through the data taking period
presented in this Letter. This reduces the radon induced background for
this analysis by about a factor of six.
Other external background contributions are found to be small.

The background activities are determined by measuring
control decay channels that are independent of the signal topology
using a full MC simulation of the background processes and the detector. 
A summary of the measured background activities are given in Table~\ref{table1}.

In the $e\gamma$ channel, $^{207}$Pb from the $^{207}$Bi decays
via the emission of an electron and a photon 
from the strongly-converted energy transition in excited $^{207}$Pb. 
$^{152}$Eu decays into excited  $^{152}$Gd through $\beta$ decay 
which de-excites into the ground state via photon emission.
\begin{figure}[htb] 
\begin{center} 
\includegraphics[width=0.35\textwidth]{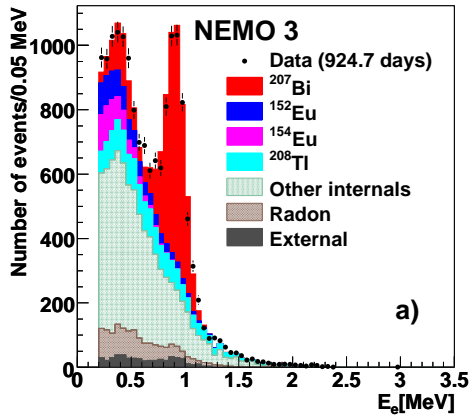}
\includegraphics[width=0.35\textwidth]{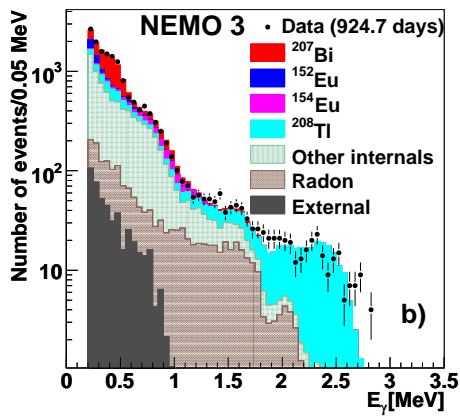}
\end{center} 
\caption[]{(color online) Energy distributions of the a) electrons and b) photons in 
$e\gamma$ events.
The data with statistical uncertainties 
are compared to the sum of the expected background.
}
\label{fig-egamma} 
\end{figure} 
The $e\gamma$ events are selected by requiring that 
exactly one negatively charged particle with a track length greater than 50~cm
is found in the sector containing the $^{150}$Nd foil. The track must be
associated with an isolated scintillator
hit with energy greater than 0.2~MeV. It must originate
from the $^{150}$Nd foil and must have a hit in the first layer 
of the Geiger cells.
The scintillator measurement of the time-of-flight of the electron candidates
has to be consistent with the hypothesis that the decay occurred in the source
foil. The photon is identified by requiring that there is a second
scintillator hit with no associated track and with energy greater than 0.2~MeV.
The energy sum of the all other clusters not associated to
the electron or the photon  must be less than 0.15~MeV.
The opening angle between the electron 
and the photon is required to be $\cos\theta<0.9$.

To obtain the background normalisation for $^{207}$Bi and
$^{152}$Eu, the normalization of these background contributions is fitted to spectra
of the electron energies, $E_e$, and the photon energies, $E_{\gamma}$, 
shown in Figure~\ref{fig-egamma}.
The contributions from other background sources are fixed in the
fit. The $^{228}$Ac contribution is taken from the $^{208}$Tl component
in the high energy tail of the $E_{\gamma}$ distribution, since 
$^{208}$Tl originates from $^{228}$Ac decays.
The $^{212}$Bi component is normalised with respect to the measurement
of its decay product $^{208}$Tl. 
The external, the radon induced, $^{214}$Bi and $^{214}$Pb 
backgrounds are set to their independently measured values. 

\begin{figure}[htb] 
\begin{center} 
\includegraphics[width=0.35\textwidth]{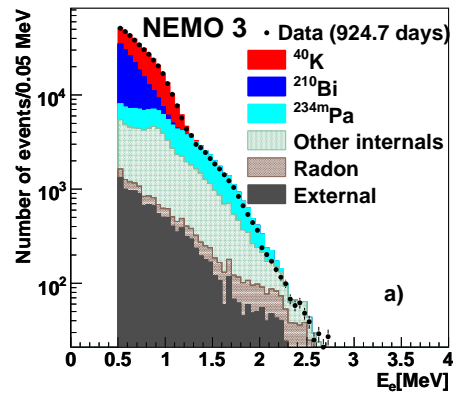}
\end{center} 
\caption[]{(color online) Distribution of the electron energy in single electron events.
The data are compared to the sum of the expected background. 
}
\label{fig-single} 
\end{figure} 

The isotopes $^{234m}$Pa and $^{40}$K undergo $\beta$ decay.
Their activities are therefore measured with
single electron events. The same selection for the electron 
is applied as in the $e\gamma$ channel, apart from an additional
requirement $E_e>0.5$~MeV. The $^{234m}$Pa and $^{40}$K contributions
are obtained from a fit of the normalization to the $E_e$ distributions. 
The external background is set to the values obtained from
independent measurements. The $^{207}$Bi
component has been measured in the $e\gamma$ channel.
The $E_e$ distribution with
the fitted $^{234m}$Pa and $^{40}$K contributions is shown in
Figure~\ref{fig-single}.
The distribution of the event vertices for the single
electron events, calculated as the
electron track's intersection with the source foil, 
shows small regions of high activity consistent with $^{234m}$Pa contamination
which are removed from the analysis. 
The activities given in Table~\ref{table1}.
are measured after the removal of these regions.

Due to the resolution of the tracking detector, events from neighboring 
source foils ($^{48}$Ca, $^{96}$Zr, $^{100}$Mo) 
can be reconstructed  as originating from the $^{150}$Nd foil. The 
total number of background events from neighboring foils is estimated
to be $126\pm12$ from MC simulations, where the uncertainty is dominated
by the uncertainty on the half-life of these isotopes.

\begin{table}[htb]
\begin{center}
    \begin{tabular}{|c|c|c|c|}
      \hline
     Background & $A$~(mBq/kg) &  $N_{bg}$ &  $N_{bg}$~for         \\
                 &              &            &$\Esum>2.5$~MeV      \\
      \hline 
      $^{152}$Eu  &  $54\pm 6$ &  $\phantom{0}19 \pm 1 \phantom{0}  $  & $0$\\
      $^{154}$Eu  &  $22\pm 2$ &  $\phantom{0}9 \pm 1 $  & $0$\\
      $^{208}$Tl  &  $10\pm 2$  &  $\phantom{0}46 \pm 4\phantom{0}$& $3.5 \pm 0.9$ \\
      $^{228}$Ac  &  $27.8 \pm 5.6$  & $\phantom{0}52 \pm 5\phantom{0}$         & $0$\\
      $^{212}$Bi  &  $27.8 \pm 5.6$  & $\phantom{0}32 \pm 4\phantom{0}$         & $0$\\
      $^{207}$Bi  &  $231 \pm 10 $ & $\phantom{0}138 \pm 6\phantom{00}$ & $0$      \\
      $^{214}$Bi  &  $3.3\pm 0.8$  & $\phantom{0}13 \pm 3\phantom{0}$ & $0.6\pm0.2$\\
       $^{214}$Pb  &  $3.3\pm 0.8$  & $\phantom{0}6 \pm 1$ & $0$\\
      $^{40}$K    & $213 \pm 10 $& $ \phantom{0}66 \pm 6 \phantom{0} $      &  $0$ \\
      $^{234m}$Pa &  $47 \pm 2$     & $\phantom{0}143 \pm 6\phantom{00}$  & $0$  \\
       External and  &    &     &   \\
       radon induced &    &   $\phantom{0}53 \pm 11$  & $4.8\pm0.8$  \\
$^{48}$Ca, $^{96}$Zr, $^{100}$Mo  
&              &     $\phantom{0}168 \pm 19\phantom{0}$            &  $0.12\pm 0.02$   \\
\hline
     sum & &$\phantom{0}746 \pm 30\phantom{0}$&  $9.0 \pm 1.2$ \\
\hline
    data  & & $2789$  &  $29$ \\
\hline
   \end{tabular}
      \caption{Summary of the measured background activities,
the expected number of background events and the
observed number of events for the selected data set
and in the high energy region, $\Esum>2.5$~MeV. }
\label{table1}
  \end{center}
\end{table}

\begin{figure}[htb] 
\begin{center} 
\includegraphics[width=0.35\textwidth]{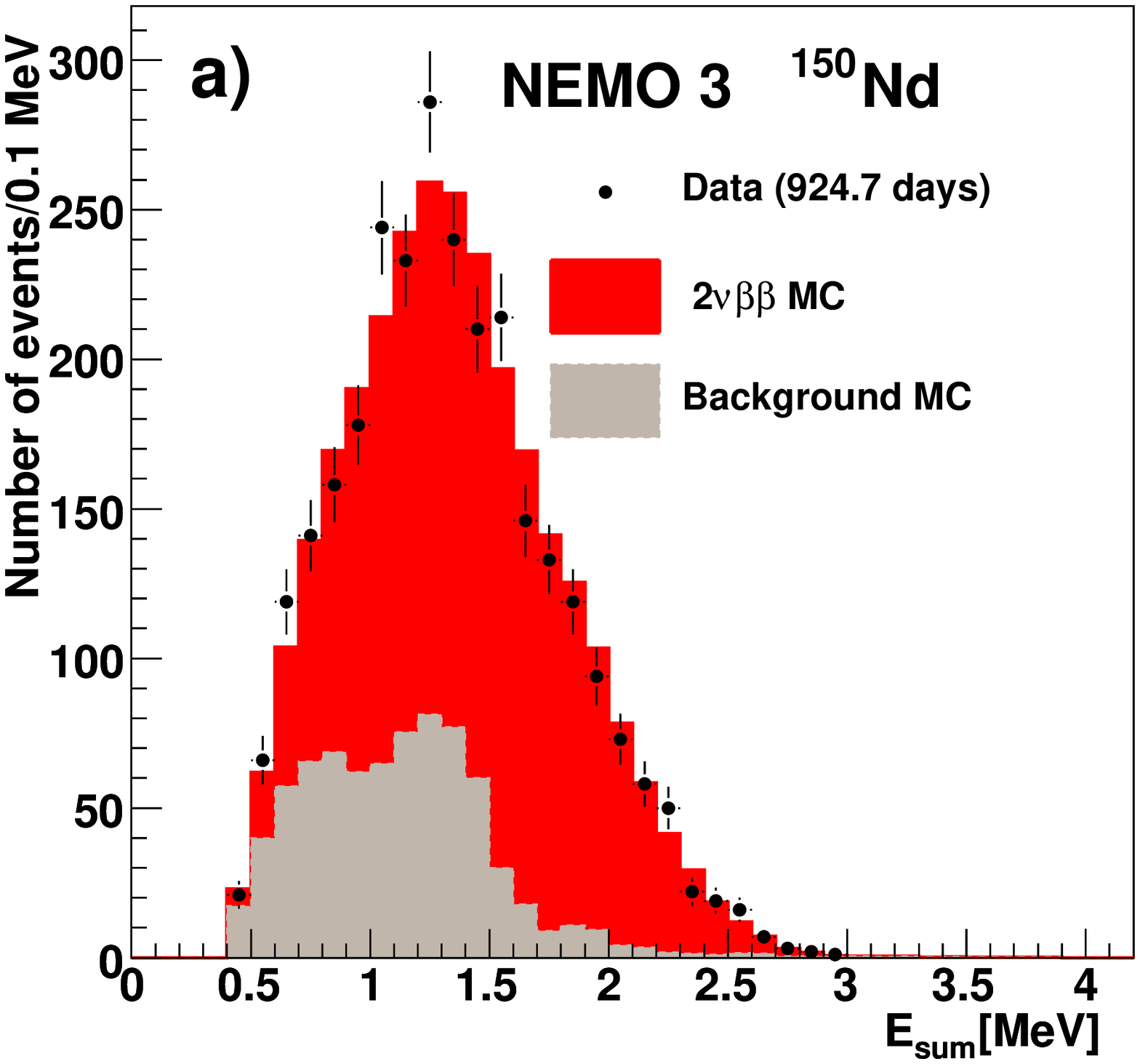}
\includegraphics[width=0.35\textwidth]{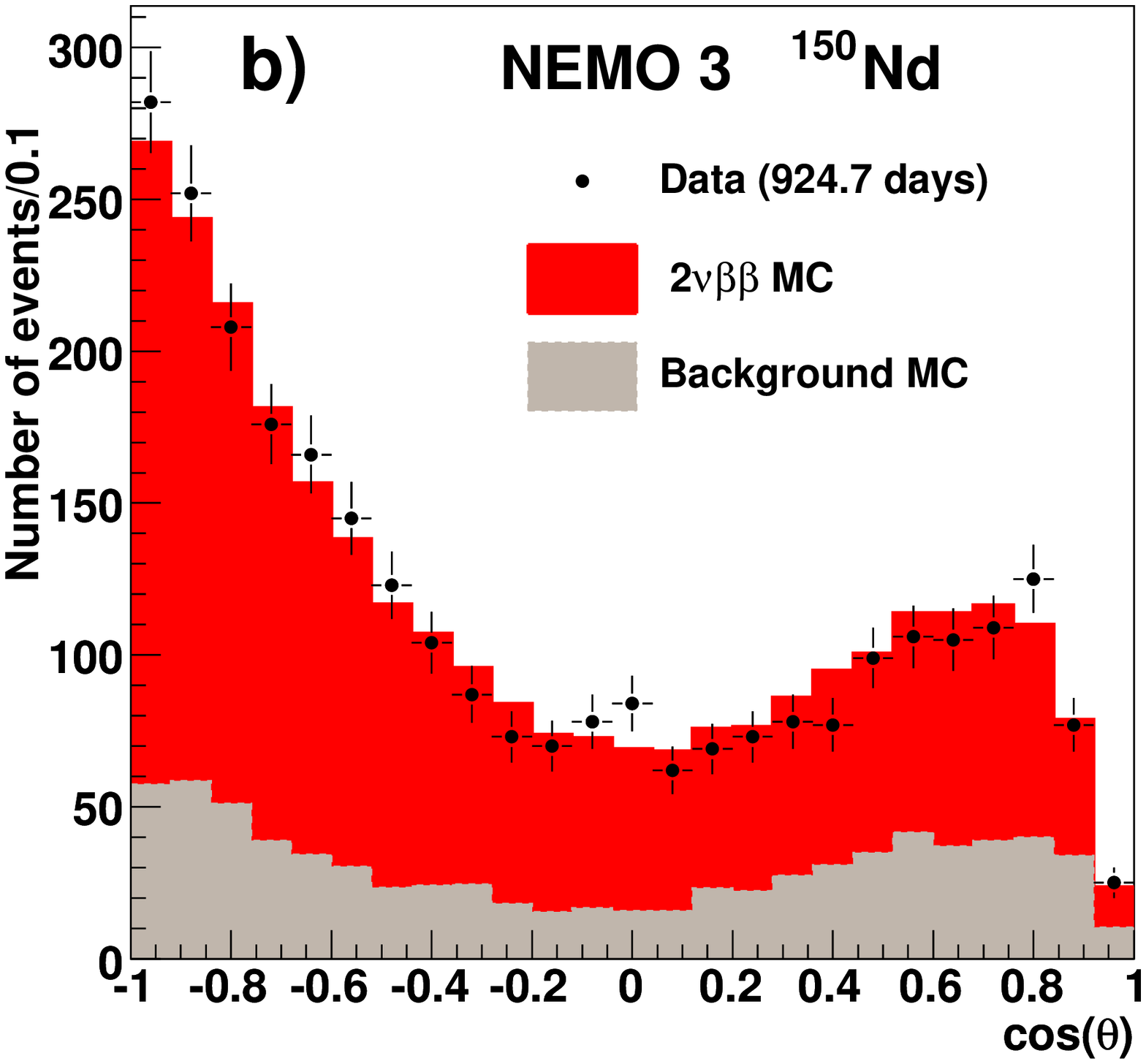}
\end{center} 
\caption[]{(color online) Distributions of  a) the energy sum of the two electrons, $\Esum$,
and b) the angle between the two electrons, $\cos\theta$,
for data compared to the sum of the background and $2\nu\beta\beta$
signal expectations.
}
\label{fig-esum} 
\end{figure} 

The distributions of the energy sum of the two electrons and
the opening angle between them are shown in Figure~\ref{fig-esum}.
The data are in good agreement with the sum of the background
and the $2\nu\beta\beta$ signal distributions. We therefore
use the data sample to measure the $2\nu\beta\beta$ half-life, $T_{1/2}^{2\nu}$,
of $^{150}$Nd. The efficiency of the $2\nu\beta\beta$ event selection 
is $7.2\%$. After background subtraction we obtain
\begin{equation}
  T_{1/2}^{2\nu} = (9.11^{+0.25}_{-0.22}(\rm stat.) \pm 0.63~(\rm syst.)) 
\times 10^{18}~\rm y\, .
 \end{equation}

The systematic uncertainty on the sum
of the internal and external background is $4.3\%$ which
translates into an uncertainty on the $2\nu\beta\beta$
half-life of $1.6\%$. 
This includes the uncertainty 
from the background fits and from the measurement of
the activity of $^{208}$Tl using two different decay channels.
The uncertainty on the tracking efficiency is $5.7\%$.
Varying the TOF requirement leads to a $1\%$ uncertainty. 
An uncertainty of $3\%$ is due to the uncertainty on the 
position of the $^{150}$Nd foil in the detector.
The uncertainty on the energy measurement was studied by smearing the 
individual electron
energies by $2\%$ which yields a systematic uncertainty
of $1.5\%$ on the $2\nu\beta\beta$ half-life. 

Since no significant excess is observed in the $\Esum$ distribution, 
a limit is set on the half-life for neutrinoless double beta 
decay, $T_{1/2}^{0\nu}$, using the $CL_s$ method~\cite{bib-cls}. 
Only the shapes and not the normalization of the full $\Esum$ distribution
are used to discriminate signal and $2\nu\beta\beta$ background.
The normalization of the $\Esum$ distribution for
the other backgrounds is fixed to the value given in Table~\ref{table1}.
All limits in this Letter are calculated by utilizing a likelihood-fitter~\cite{bib-wade} 
that uses a log-likelihood ratio (LLR) test statistic method.
Two hypotheses are defined, the signal-plus-background hypothesis and 
the background-only hypothesis. The LLR distributions
are populated using Poisson simulations of the two hypotheses.
Systematic uncertainties are treated as uncertainties on the expected 
numbers of events and are folded into the signal and 
background expectations via a Gaussian distribution.
Correlations between systematic uncertainties are taken
into account.
The value of the confidence level, $CL_s$, is defined as 
$CL_s = CL_{s+b} / CL_b$,
where $CL_{s+b}$ and $CL_b$~are the confidence levels in the 
signal-plus-background and background only hypotheses, respectively.
The limits are calculated by scaling the signal until $1-CL_s$ reaches $0.9$.

\begin{figure}[htb] 
\begin{center} 
\includegraphics[width=0.35\textwidth]{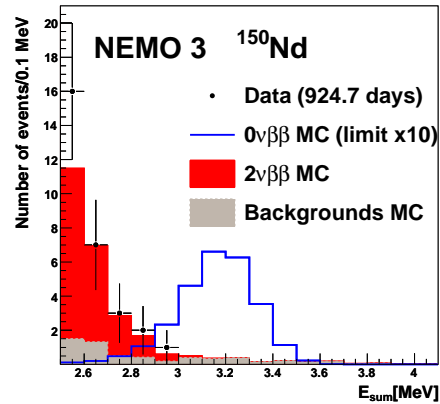}
\end{center} 
\caption[]{(color online) Distribution of the energy sum of the 
two electrons, $\Esum$, for $\Esum>2.5$~MeV. The
data are compared to the total background, consisting
of internal and external background and the 
$2\nu\beta\beta$ expectation.
A MC simulation of a $0\nu\beta\beta$ signal with a
half-life of $1.8 \times 10^{21}$~years, corresponding to
ten times the number of events expected for
the observed $90\%$ C.L. limit, is also shown.
}
\label{fig-signal} 
\end{figure} 

The total efficiency for $0\nu\beta\beta$ events after applying
all selections is $(19 \pm 1)\%$. The uncertainties on the efficiency
of the signal and the background are assumed to be fully correlated.
The $\Esum$ distribution for $\Esum>2.5$~MeV is shown in 
Figure~\ref{fig-signal} for data compared to the total 
background, which consists 
of internal and external background and the $2\nu\beta\beta$ expectation.
A MC simulation of a $0\nu\beta\beta$ signal is also shown.
Signal and background are well separated, demonstrating
the advantages of using $^{150}$Nd for future $0\nu\beta\beta$ searches.
The observed limit on the half-life is
$T_{1/2}^{0\nu}>1.8 \times 10^{22}$~years
at $90\%$ Confidence Level (CL). It is consistent with the 
median expected limit.

This limit on the half-life is converted into a limit on
the effective Majorana neutrino mass, $\langle m_{\nu} \rangle$, 
using nuclear matrix elements (NME).
In the currently available 
QRPA-like calculations spherical symmetry of the nucleus
has been assumed. If the upper and lower limits on the calculated NME, 
which also include the uncertainties in the weak coupling constant $g_A$, 
are taken into account~\cite{bib-QRPA}, the experimental lower limit on 
the half-life of $^{150}$Nd translates into a limit 
$\langle m_\nu\rangle < 1.5 -2.5$~eV. 

Taking into account the nuclear deformation will modify this conclusion. 
The suppression of the NME for $^{150}$Nd has been estimated to be a
factor $2.7$ in the case of nuclear deformations derived from laboratory
moments~\cite{defac}. This increases the upper limit to
$\langle m_{\nu} \rangle < 4.0 -6.8$~eV which is 
consistent with the limit derived using the NME of a pseudo SU(3) 
model~\cite{bib-SU3} and the PHFB model~\cite{bib-PHFB}, which also
include the effect of nuclear deformation. Further progress
in the calculation of the NME for $0\nu\beta\beta$ decay of $^{150}$Nd 
is therefore urgently required.

Neutrinoless double beta decay can also proceed through the emission
of one or two Majorons ($\chi^0$)~\cite{bib-bamert}.
These decays are characterized by the spectral index $n$ which
leads to a modification of the energy sum distribution by a factor
$(Q_{\beta\beta}-\Esum)^n$. We consider models with $n=1,3,7$~\cite{bib-bamert}
and $n=2$~\cite{bib-bulk}. Limits on $n=3,7$ models are less sensitive
than $n=1,2$ since the shape of the $\Esum$ distribution is
similar to the one for $2\nu\beta\beta$ decay ($n=5$).
The $CL_s$ method is applied to the $\Esum$ distribution
in the same way as described previously. Since the normalisation
of the $2\nu\beta\beta$ background can not be independently
determined, it is left unconstrained in the fit determining the limits.

We also set limits for $0\nu\beta\beta$
processes involving right-handed currents (V+A). 
In addition, 
the $0\nu\beta\beta$ decay can also proceed through various excited
states ($0_1^+$ or $2_1^+$). Limits on all these modes of 
neutrinoless double beta decay are shown in Table~\ref{table-others}.
They either improve previously published limits~\cite{bib-silva,bib-prev} 
significantly or represent the first limits for this isotope ($n=2,3,7$).
The limit on $n=1$ Majoron emission can be be translated into
a limit on the neutrino-Majoron coupling of $g_{ee}<(0.64-1.05) \times 10^{-4}$
without and $g_{ee}<(1.7-3.0) \times 10^{-4}$ with nuclear deformation. 
These limits
are comparable with the limits obtained for $^{82}$Se and $^{100}$Mo
with about 9 and 70 times the exposure, defined by isotope mass times observation
time~\cite{bib-momaj}.

\begin{table}[htb]
 \begin{center}
    \begin{tabular}{|c|c|c|c|c|c|c|c|c|}  
    \hline
 & \multicolumn{4}{c|}{$0\nu\beta\beta$} & 
   \multicolumn{4}{c|} {Majorons}           \\
\cline{2-9} 
 & $0^{+}_{\rm gs}$ ($\langle m_{\nu}\rangle$) 
 & $0^{+}_{\rm gs}$(V+A) 
 & $2^{+}_{1}$  
 & $0^{+}_{1}$  
 & $n=1$ &  $n=2$ & $n=3$ & $n=7$   \\
\cline{1-9} 
limit      & $18.0$  &$10.7$ & $2.4$  & $0.24$  & $1.52$  & $0.54$ & $0.22$  &$0.047$\\
\cline{1-9}
\hline
      \end{tabular}
      \caption{$90\%$ C.L. 
limits on the half-life, $T_{1/2}$, in units of $10^{21}$~y,  
for different modes of neutrinoless double beta decay.} 
     \label{table-others}
  \end{center}
\end{table}

In summary, we have presented the most precise measurement
of the half-life of double beta decay of $^{150}$Nd to date, 
yielding a value of 
$  T_{1/2}^{2\nu} = (9.11^{+0.25}_{-0.22}(\rm stat.) \pm 0.63~(\rm syst.)) 
\times 10^{18}$~years. This value is slightly more than two
standard deviations higher than the previously measured value
$  T_{1/2}^{2\nu} = (6.75^{+0.37}_{-0.42}(\rm stat.) \pm 0.68~(\rm syst.)) 
\times 10^{18}$~years~\cite{bib-silva}. 
We have  significantly
improved limits on the half-life of different modes of neutrinoless double 
beta decay for $^{150}$Nd and we have provided the first limits
for several Majoron models using $^{150}$Nd. 
Our measurements demonstrate that $^{150}$Nd is
an excellent candidate isotope for future double beta decay experiments.

We thank the staff at the Modane Underground Laboratory 
for its technical assistance in running the experiment, Vladimir
Tretyak for providing the Monte Carlo event generator and
Wade Fisher for helping with the limit setting program..
We acknowledge support by the Grants Agencies 
of the Czech Republic, RFBR (Russia), STFC (UK) and NSF (USA).

\end{document}